\documentclass[twocolumn,trackchanges]{aastex62}
\usepackage{url,amssymb,amsmath,rotating,color,units,wasysym,epsfig,multirow,epstopdf}

\usepackage{graphicx}% Include figure files
\usepackage{dcolumn}% Align table columns on decimal point
\usepackage{bm}% bold math

\newcommand{\Pdet}{P_{\text{det}}}
\newcommand{\pdet}{p_{\text{det}}}

\newcommand{\comment}[1]{}
\newcommand{\revision}[1]{#1}

\newcommand{\Caltech}{LIGO Laboratory, California Institute of Technology, Pasadena, CA 91125, USA}
\newcommand{\Monash}{School of Physics and Astronomy, Monash University, VIC 3800, Australia}
\newcommand{\OzGrav}{OzGrav: The ARC Centre of Excellence for Gravitational-Wave Discovery, Clayton, VIC 3800, Australia}

\begin{document}

\title{Flexible and accurate evaluation of gravitational-wave Malmquist bias with machine learning}

\author[0000-0003-2053-5582]{Colm~Talbot}
\email{colm.talbot@ligo.org}
\affiliation{LIGO Laboratory, Massachusetts Institute of Technology, Cambridge, Massachusetts 02139, USA}
\affiliation{Kavli Institute for Astrophysics and Space Research, Massachusetts Institute of Technology, Cambridge, Massachusetts 02139, USA}
\affiliation{\Caltech}

\author[0000-0002-4418-3895]{Eric~Thrane}
\affiliation{\Monash}
\affiliation{\OzGrav}

\date{\today}

\begin{abstract}
Many astronomical surveys are limited by the brightness of the sources, and gravitational-wave searches are no exception.
The detectability of gravitational waves from merging binaries is affected by the mass and spin of the constituent compact objects.
To perform unbiased inference on the distribution of compact binaries, it is necessary to account for this selection effect, which is known as Malmquist bias.
Since systematic error from selection effects grows with the number of events, it will be increasingly important over the coming years to accurately estimate the observational selection function for gravitational-wave astronomy.
We employ \revision{density estimation} methods to accurately and efficiently compute the compact binary coalescence selection function.
We introduce a simple pre-processing method, which significantly reduces the complexity of the required machine learning models.
\revision{We demonstrate that our method has smaller statistical errors at comparable computational cost than the method currently most widely used allowing us to probe narrower distributions of spin magnitudes.
The currently used method leaves $10-50\%$ of the interesting black hole spin models inaccessible; our new method can probe $>99\%$ of the models and has a lower uncertainty for $>80\%$ of the models.
}

\end{abstract}

\section{\label{sec:motivation}Motivation}
The phenomenon of Malmquist selection biases in brightness-limited astronomical surveys has been known for nearly a century~\citep{Malmquist1922, Malmquist1925}, and methods for mitigating this bias have remained an active area of research in astronomy (e.g.,~\cite{Freudling1995, Loredo2004, Farr2015, ForemanMackey2014, March2018, Connor2019, Mandel2019}).
Gravitational-wave searches are affected by Malmquist bias as the parameters of the merging binaries determine the luminosity of the source.
More massive binaries produce larger gravitational-wave strain, all else equal, and so they can be observed at greater distances, at least until the mass becomes so large that the signal begins to shift out of the observing band.
Additionally, binaries with black holes that are spinning \revision{along axes} aligned \revision{with} the \revision{orbital} angular momentum are visible to further distances~\citep{Campanelli2006, Ng2018}.

Over the past several years, there has been increasing interest in population studies, which seek to measure the distribution of astrophysical parameters such as the mass, spin, and distance of merging compact objects using events observed with Advanced LIGO/Virgo~\citep{aLIGO, AdVirgo} (see, e.g.,~\cite{O3aPop, Roulet2021, O3bPop} and references therein).
To perform unbiased inference on the distribution of astrophysical parameters, it is necessary to account for selection biases when performing population inference, see, e.g.,~\cite{Loredo2004, Farr2015,Mandel2019,Thrane2019,Vitale2020}.
The standard method employed in gravitational-wave searches requires computing the \revision{total sensitivity of search pipelines} for \revision{a given} population model.
Evaluating the \revision{sensitivity} for different population parameters involves integrating the \revision{time-dependent sensitivity to} all binary parameters over the whole observing time.

Astrophysical population inference is \revision{typically} performed using Markov Chain Monte Carlo (MCMC)\, \citep{Metropolis1953, Hastings1970} or nested sampling algorithms~\citep{Skilling04}, which \revision{can require $O(10^3-10^7)$} likelihood evaluations for a well converged run, and \revision{the selection function must be} evaluated on the fly at every iteration.
As the binary black hole catalog grows this integral must be evaluated with increasing precision~\citep{Farr2019}, and correspondingly increased computational cost.
Additionally, as the catalog grows, so does our resolving power, meaning that sub-dominant effects, e.g., the effect of black hole spin on the \revision{sensitivity}, must be considered.
\revision{Guaranteeing sufficient precision is especially challenging for narrow population models as the Monte Carlo integrals currently performed are poorly suited to probing these distributions.
In this work, we demonstrate that by performing a density estimation step on the set of found injections, we can dramatically increase the efficiency of these calculations, enabling us to probe narrow population models.
}

The rest of this paper is structured as follows.
In the next section, we \revision{define some relevant quantities and} outline methods for accounting for Malmquist bias in gravitational-wave searches.
We then briefly summarize a few preliminaries for gravitational-wave population inference in Section~\ref{sec:popinf}.
Following this, in Section~\ref{sec:density}, we describe the problem of density estimation and discuss various commonly used methods.
In Section~\ref{sec:implementation} we estimate the gravitational-wave transient selection function as a function of binary parameters using a Gaussian mixture model.
After this, we apply our methods to the binary black hole systems identified in~\cite{GWTC3}.
Some closing thoughts are then provided.
The notebook that performed the analysis presented here can be found at~\cite{notebook}.

\section{\label{sec:vt}\revision{Sensitivity} estimation}
Most gravitational-wave population analyses impose a detection threshold on the analyzed triggers to avoid contamination from terrestrial noise sources\revision{, for example, demanding that the false-alarm rate of a trigger is less than once per year}.
Applying this threshold leads to a selection bias in the observed sample.
We quantify this by considering the probability that a signal with parameters $\theta$ \revision{(e.g., binary mass and black hole spin)} would surpass our threshold $\rho_{th}$\footnote{The specific choice of threshold is irrelevant so long as it is robustly defined.}
\begin{equation}
    p_{\text{det}}(\theta)
    = \int_{\rho > \rho_{\text{th}}} d{\bf d} \, p({\bf d}|\theta).
\label{eq:pdet}
\end{equation}
The integral is over all observed data \revision{and $\pdet$ is the fraction of the data which surpasses the threshold under the assumption that a signal with parameters $\theta$ is present}.
For population analyses, we require the fraction of all sources that are detectable, for a given population model, characterized by parameters, $\Lambda$,
\begin{equation}
        P_{\text{det}}(\Lambda) = \int \, d\theta \, p(\theta | \Lambda) \, p_{\text{det}}(\theta),
\label{eq:Pdet}
\end{equation}
where $p(\theta|\Lambda)$ is a conditional prior for $\theta$ given population (hyper-) parameters $\Lambda$, e.g., the shape of the black hole mass distribution.
For a detailed derivation of these quantities see, e.g.,~\cite{Finn1993, Messenger2013, Farr2015, Tiwari2018, Thrane2019, Mandel2019}.

We emphasize that all population analyses which apply a threshold necessarily have a corresponding selection bias that must be accounted for, including analyses that explicitly model contamination of the sample from terrestrial sources~\citep{Gaebel2019, Galaudage2020, Roulet2020}.
However, see~\cite{TBS-PE} for a method that avoids thresholds entirely.

The integral over ${\bf d}$ in Eq.~\ref{eq:pdet} requires that we understand the sensitivity of gravitational-wave searches throughout the observing history.
In practice, there are currently two widely used methods to compute this integral: inject simulated signals into the data and see how many of them are recovered by the search pipelines; or use a semi-analytic approximation based on the power spectral density of the interferometers, e.g.,~\cite{Finn1993}.

The former method gives the most faithful representation of the search sensitivity.
However, the latter has several computational advantages.
Because of the large parameter space that must be covered, the injection and recovery procedure gives us only the parameter values of the found/missed signals, whereas the semi-analytic approach can efficiently generate a numerical value for $p_{\det}$ marginalized over specific nuisance parameters.
Thus, the semi-analytic approach can also be performed much more computationally cheaply due to the cost of performing and recovering injections.
Previous methods to improve the reliability of semi-analytic estimates include calibration of semi-analytic estimates with the output of injection campaigns~\citep{WysockiVT} \revision{and phenomenological fits, e.g.,~\cite{Fishbach2018, Veske2021}}.

\revision{There have been several recent methods to leverage supervised machine learning methods to estimate $\pdet$~\citep{Gerosa2020, Wong2020b}.
\citeauthor{Gerosa2020}~use a neural network classifier to give a binary outcome of detectable or not detectable for a given set of binary parameters; this method requires retraining when the threshold is changed.
In~\citeauthor{Wong2020b}, the authors train a neural network regressor to estimate the signal-to-noise ratio (a commonly used detection threshold), allowing for the threshold to be changed trivially.
However, both of these methods require specifying all of the binary parameters in order to evaluate $\pdet$.
}
In this work, we use density estimation on the set of found injections to provide a continuous, generative, model for $\pdet$ \revision{in arbitrary subsets of the binary parameters}.

The integral over $\theta$ in Eq.~\ref{eq:Pdet} marginalizes over all the parameters describing the source---15 parameters to completely characterize a quasi-circular binary black hole merger---in addition to any parameters describing the state of the instruments.
In practice, many of the parameters are not modeled in current population analyses; the most complex models considered currently fit for the distribution of seven of these parameters, the two component masses, spin magnitudes, spin-tilt angles, and redshift, requiring the evaluation of a seven-dimensional integral within each likelihood evaluation.
The other parameters are assumed to be well described by the prior distributions used during sampling.
These are mostly geometric parameters describing the position and orientation of the binary, although it is possible that some of these parameters may deviate from isotropy.
For example, we could search for deviations from isotropy over the sky position, e.g.~\cite{Payne2020, Stiskalek2021}, or features in the distribution of the azimuthal spin parameters due be influenced by spin-orbit responances~\citep{Schnittman2004, Gerosa2018b, Varma2021}.
This integral is, therefore, recast as a Monte Carlo integral over the set of found injections~\citep{Tiwari2018, Farr2019}
\begin{equation}
    P_{\text{det}}(\Lambda) = \frac{1}{N_{\text{inj}}} \sum_{i=1}^{N_{\text{found}}} \frac{p(\theta_i | \Lambda)}{p(\theta_i | \Lambda_0)}.
\label{eq:Pdetmc}
\end{equation}
Here, $p(\theta|\Lambda_0)$ is the distribution of the injected signals \revision{which will depend on the specific analysis}, $N_{\text{inj}}$ is the total number of injected signals and $N_{\text{found}}$ is the number of injections \revision{surpassing the threshold}.
\revision{The sum in Equation~\ref{eq:Pdetmc} is over samples drawn from the distribution of found injections
\begin{equation}
    \theta_{i} \sim \pdet(\theta_{i}) \pi(\theta_{i} | \Lambda_{0}).
\end{equation}}
To ensure sufficient convergence of the Monte Carlo integral we must have an effective sample size of at least four times the number of observed events~\citep{Farr2019}.
This means that to fit tightly peaked distributions we need a large number of samples for the distribution of found injections or a continuous representation of $p_{\text{det}}$.
Performing more injections to increase the number of recovered injections quickly becomes computationally prohibitive.
In this work, we resolve this issue by performing density estimation using the set of found injections.
Using these density estimates, we can directly evaluate $\pdet$ and/or generate additional samples from the distribution of found injections.

\section{\label{sec:popinf}Population inference}

\subsection{\label{sec:models}Models}

For demonstration purposes, we consider two simple population models from within the gravitational-wave literature.
Following~\cite{O3aPop, Talbot2018a}, we model the binary black hole mass distribution as a power law in the larger mass, $m_1$, between the minimum and maximum mass along with a normally distributed component and a power law in the mass ratio, $q = m_2 / m_1$,
\begin{align}
    p(m_1 &| \alpha, m_{\min}, m_{\max}) = (1 - \lambda) \frac{(1-\alpha) m_1^{-\alpha}}{m_{\max}^{1-\alpha} - m_{\min}^{1-\alpha}} \nonumber\\
    &\quad + \frac{\lambda}{\sqrt{2\pi\sigma^2_m}} \exp\left( - \frac{(m_1 - \mu_m)^2}{2\sigma^2_m} \right) \\
    p(q &| m_1, \beta, m_{\min}) =  \frac{(1+\beta) q^{\beta}}{1 - \left(\frac{m_{\min}}{m_1}\right)^{1+\beta}}\label{eq:mass_ratio}.
\end{align}
This is the \textsc{Power-Law + Peak} model in~\cite{O3aPop} without the low-mass smoothing.
We assume that both component spins are drawn from the same distribution.
We model the spin magnitudes as following a Beta distribution~\citep{Wysocki2019}
\begin{equation}
    p(a_i | \alpha_{\chi}, \beta_{\chi}) = \frac{a_i^{\alpha_{\chi} - 1} (1 - a_i)^{\alpha_{\chi} - 1}}{\mathrm{B}(\alpha_{\chi}, \beta_{\chi})}.
\end{equation}
We model the distribution of spin orientations as a combination of a truncated half-normal and a uniform distribution~\citep{Talbot2017}
\begin{equation}
    p(\cos\theta_i | \sigma_{s}) = \frac{(1 - \xi)}{2} + \xi N(\sigma_s) \exp\left( - \frac{(\cos\theta_i - 1)^2}{2 \sigma^{2}_{s}} \right)
\end{equation}
The factor $N$ ensures that the distribution is properly normalized.
This is the \textsc{Default} model in \cite{O3aPop}.

The reference distribution using the LIGO/Virgo/Kagra collaborations' most recent injection campaign~\citep{O3bSensitivity} is the product of these distributions
with population hyper-parameters $\alpha=2.35$, $m_{\min}=2$, $m_{\max}=100$, $\lambda=0$, $\beta=1$, \revision{$\alpha_\chi=1$, $\beta_\chi=1$, and $\xi=0$. 
These define $p(\theta|\Lambda_0)$ for our application}.

\subsection{\label{sec:poplike}Likelihood}

The standard likelihood used in population inference for gravitational-wave sources in the presence of selection biases is (e.g.,~\cite{Thrane2019, Mandel2019, Vitale2020}),
\begin{equation}
    \mathcal{L}(\{d_i\}|\Lambda) = \frac{1}{P_{\text{det}}(\Lambda)^N} \prod^{N}_{i} \int d\theta_i \mathcal{L}(d_i | \theta_i) p(\theta_i | \Lambda).
\label{eq:likelihood}
\end{equation}
Where the product over $i$ runs over the $N$ observed events with data $d_i$.
The integral over $\theta_i$ is typically performed by importance sampling from the single-event posterior distribution for $p(\theta_i | d_i)$ as is done to calculate $P_{\text{det}}$.
We take the publicly available samples from the single-event posterior distributions from~\cite{GWTC1, GWTC2, GWTC3}.
\revision{Since the likelihood explicitly depends on $\Pdet$, calculating this quantity is the main target of this work.
This likelihood is then used to explore the posterior distribution for the population parameters given all of the observed data, e.g., in Section~\ref{sec:hyperpe}.}

\section{\label{sec:density}Density estimation}

Reconstructing a function or probability density from a finite set of samples from the distribution is a widespread problem in data analysis.
For example, injection campaigns to determine the sensitivity of gravitational-wave detectors do not give us a continuous description of the sensitivity, but rather a discrete set of samples from the distribution of found injections.

Density estimation methods can be loosely divided into parametric and non-parametric methods.\footnote{We note that the word ``non-parametric'' is something of a misnomer, as often these models involve large numbers of unphysical parameters to perform the fit. An alternative delineation is between models where the parameters are either physically motivated (parametric) or not physically motivated (non-parametric).}
Parametric density estimation involves fitting a parameterized phenomenological model to the data.
An example of this is the method used to reconstruct the population distribution of binary black holes in this work.
To estimate the gravitational-wave selection function we will rely on non-parametric density estimation.

Many methods for nonparametric density estimation are commonly used; however, most traditional methods such as binning or kernel density estimation scale poorly as the dimensionality of the problem increases.
More sophisticated density estimation techniques involving the optimization of many parameters, such as Gaussian mixture models or flow-based inference, have proved successful at approximating complex functions in large dimensional spaces, see, e.g.,~\cite{Powell2019, Gabbard2019, Green2020a, Green2021, Wong2020a, Wong2020b, Wong2021a} for applications in gravitational-wave inference.
These models also provide natural ways to generate additional samples from the underlying densities \revision{at minimal cost} and are therefore sometimes referred to as generative models.

In this work, we approximate $\pdet(\theta)$ using a Gaussian mixture model.
A Gaussian mixture model is an unsupervised density estimator that approximates the distribution as a set of multivariate Gaussian distributions each with a unique mean and covariance.
\revision{
The model assumes that the target distribution can be well modeled by a finite sum of multivariate Gaussian distributions
\begin{equation}
    {\cal D}(\theta') = \sum^{K}_{k=1} \frac{1}{w_{k}} {\cal N}(\theta'; \mu_{k}, \Sigma_{k}).
\end{equation}
Here $K$ is the number of components in the mixture and can be manually tuned, $w_{k}$ is the weight associated with the $k$th component, $\mu_{k}$ and $\Sigma_{k}$ are the mean vector and covariance matrix for that component.
The values of the $w_{k}$, $\mu_{k}$, and $\Sigma_{k}$ are optimized using the expectation-maximization algorithm~\citep{Webb1999} to maximize the value of
\begin{equation}
    \left< \ln {\cal D} \right> = \frac{1}{N} \sum^{N}_{i=1} \ln {\cal D}(\theta'_{i})
\end{equation}
over the training data.
}

\revision{The non-parametric methods discussed above are all examples of unsupervised learning techniques as they only do not require estimates of the target density as inputs.
There are also supervised non-parametric density estimation methods that require the target function to be evaluated at the training points; for example, Gaussian process regression and neural network regression have also been applied widely in gravitational-wave data analysis, see, e.g.~\cite{Graff2012, Veitch2015, Moore2016, Doctor2017, Lange2018, Landry2019, Williams2020, Gerosa2020, dEmilio2021}.
}

\section{\label{sec:implementation}Sensitivity to binaries}

In this section, we develop methods to evaluate Equations~\ref{eq:Pdet} and~\ref{eq:pdet} and generate new samples from the distribution of found injections.
We begin by training a function to estimate $p_{\text{det}}(\theta) p(\theta | \Lambda_0)$ using a set of $\sim 8\times 10^4$ found injections.
\revision{We use the same sensitivity data products used in~\cite{O3bPop}.
Specifically, we take the found injections from Advanced LIGO/Virgo's third observing run with a threshold of false alarm rate $< \unit[1]{\rm{yr}^{-1}}$ in any of the search pipelines employed by the LIGO/Virgo collaboration~\citep{O3bPop, GWTC3, O3bSensitivity}.
See the relevant publications and data releases for more details.}

\subsection{\label{sec:preprocessing}Pre-processing}

\begin{figure*}[p]
    \centering
    \includegraphics[width=\linewidth]{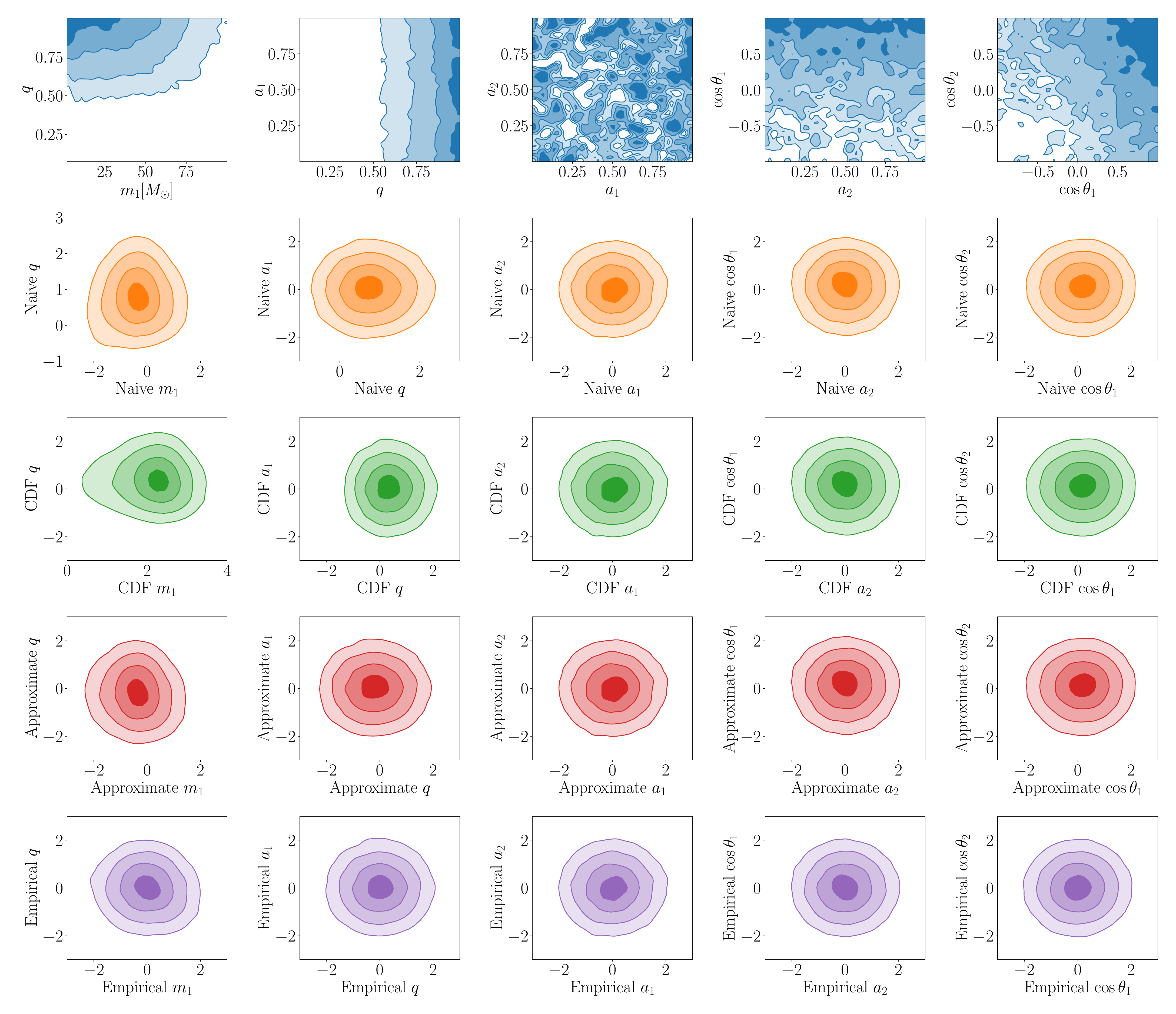}
    \caption{
    \revision{
    Two-dimensional density plots of the distribution of found injections before (top) and after (lower) applying the transformations described in Section~\ref{sec:preprocessing}.
    Our aim is to transform the data to approximate a unit multivariate normal distribution.
    In descending order these transformations are naive, CDF, approximate, and empirical (see the main text for definitions).
    We note that the empirical scaling maps the data most closely onto a unit multivariate normal distribution.
    }
    }
    \label{fig:scaling-comparison}
\end{figure*}

Gravitational-wave parameters are typically only defined over finite domains and many have significant support at the edges, e.g., spin magnitudes are contained in the unit interval and the majority of observed black holes are consistent with being non-spinning.
However, the algorithms we use for density estimation work best over an infinite domain without sharp boundaries.
Our aim is to transform the found injections such that the transformed samples are drawn from a unit multivariate normal distribution.
Therefore, we begin by performing the following mapping to the found injections:
\begin{enumerate}
    \item Transform the injections from the original distribution to the unit interval.
    \revision{We denote generic transformations as $U$ and discuss specific suitable transformations below.}
    \item Map the samples from the unit interval to a unit normal distribution using the probit function $\Phi^{-1}$~\citep{Bliss1934}.
\end{enumerate}

Mathematically the full transformation is
\begin{equation}
    \theta^{'} = \Phi^{-1}\left(U(\theta)\right)
\label{eq:regularisation}
\end{equation}
and the Jacobian is
\begin{align}
    \mathcal{J}(\theta) = \frac{d \theta'}{d\theta} = \frac{dU/d\theta}{\mathcal{N}(\theta';\mu=0,\sigma=1)}
    % \frac{p(\theta | \Lambda_0)}{\mathcal{N}(\theta';\mu=0,\sigma=1)},
\label{eq:jacobian}
\end{align}
where $\mathcal{N}$ is the normal distribution.
\revision{
We consider the four following scaling methods.

{\em Naive.}
The simplest mapping onto the unit interval is a simple shift and scale from the original domain to the unit interval
\begin{equation}
    U(\theta) = \frac{\theta - \theta_{\min}}{\theta_{\max} - \theta_{\min}}.
\end{equation}
The mapping is attractive as it can be trivially applied to any parameter and has been used in other applications, e.g.,~\cite{dEmilio2021}.
The Jacobian of this transformation is a constant
\begin{equation}
    \frac{dU}{d\theta} = \frac{1}{\theta_{\max} - \theta_{\min}}.
\end{equation}

{\em CDF.}
For some parameters the original distribution may be more complex, however, we may know an analytic form for that distribution.
In this case, we perform the mapping using the cumulative distribution function of the injected population
\begin{equation}
    U(\theta) = \int_{\theta_{\min}}^{\theta} d\bar{\theta} p(\bar{\theta} | \Lambda_0).
\end{equation}
In this case the Jacobian is
\begin{equation}
    \frac{dU}{d\theta} = p(\theta | \Lambda_{0}).
\end{equation}
This transformation is appealing as the form of the Jacobian means that evaluating $\pdet$ is trivial; however, as we will see for parameters that strongly influence the detectability, this mapping does a poor job approximating a unit uniform distribution.

{\em Approximate.}
In cases where $\pdet$ is a strong function of $\theta$ with a known (or approximately known) functional form we may choose to map onto the unit interval using an approximate expression for the observed cumulative distribution
\begin{equation}
    U(\theta) = \frac{\int_{\theta_{\min}}^{\theta} d\theta' p(\theta' | \Lambda_0) \tilde{p}_{\mathrm{det}}(\theta')}{\int_{\theta_{\min}}^{\theta_{\max}} d\theta' p(\theta' | \Lambda_0) \tilde{p}_{\mathrm{det}}(\theta')}
\end{equation}
Here $\tilde{p}_{\mathrm{det}}(\theta)$ is our analytic approximation to the selection function and the Jacobian
\begin{equation}
    \frac{dU}{d\theta} = \frac{p(\theta' | \Lambda_0) \tilde{p}_{\mathrm{det}}(\theta')}{\int_{\theta_{\min}}^{\theta_{\max}} d\theta' p(\theta' | \Lambda_0) \tilde{p}_{\mathrm{det}}(\theta')}.
\end{equation}
In our case, we use $\tilde{p}_{\rm det}(\theta) \propto m^{2.35} q^2$ so we have $p(\theta | \Lambda_0) \tilde{p}_{\mathrm{det}}(\theta) \approx {\rm const.}$
We note that the naive and approximate methods give the same transformation for primary mass and aligned spin components.
There is a difference in the transformation of the mass ratio.
While we only account for the dependence of $\pdet$ on the primary mass, we could employ a more complex expression, e.g., as discussed in~\cite{Veske2021}.

{\em Empirical.}
Finally, we use an approximation to the one-dimensional target distribution.
We construct a Gaussian kernel density estimate
\begin{align}
    \tilde{p}(\theta) &\approx p(\theta | \Lambda_{0}) p_{\rm det}(\theta) \\
    &\equiv \frac{1}{N_{\rm found}} \sum^{N_{\rm found}}_{\theta_{i}} \mathcal{N}(\theta;\mu=\theta_{i}, \sigma=\sigma)
\end{align}
from the found injections.
The standard deviation is chosen using Scott's rule~\citep{Scott1992}.
In order to account for parameters that have significant support at the edges we apply a reflecting boundary condition to the estimate
\begin{equation}
    \tilde{p}(\theta) = \tilde{p}(\theta) + \tilde{p}(2\theta_{\min} - \theta) + \tilde{p}(2\theta_{\max} - \theta)
\end{equation}
Using this estimate, we then compute an empirical cumulative distribution function
\begin{equation}
    U(\theta) = \int_{\theta_{\min}}^{\theta_{max}} d\theta' \tilde{p}(\theta').
\end{equation}
The Jacobian for this transformation is
\begin{equation}
    \frac{dU}{d\theta} = \tilde{p}(\theta)
\end{equation}
and can be trivially evaluated from our kernel density estimate.

In Figure~\ref{fig:scaling-comparison} we show the set of found injections in the original ($\theta$) space and each of the transformed spaces.
In descending order, the rows are the original data, naive scaling, CDF scaling, approximate scaling, and empirical scaling respectively.
Each of the transformations has removed the railing against the boundaries in all the parameters.
However, there are visible features remaining, especially in the mass parameters.
We note that the empirical scaling most closely transforms the data to an uncorrelated multivariate unit normal.
We will use the empirical scaling going forward unless otherwise specified.
}

\subsection{\label{sec:training}Density estimation}

\begin{figure}
    \centering
    \includegraphics[width=\linewidth]{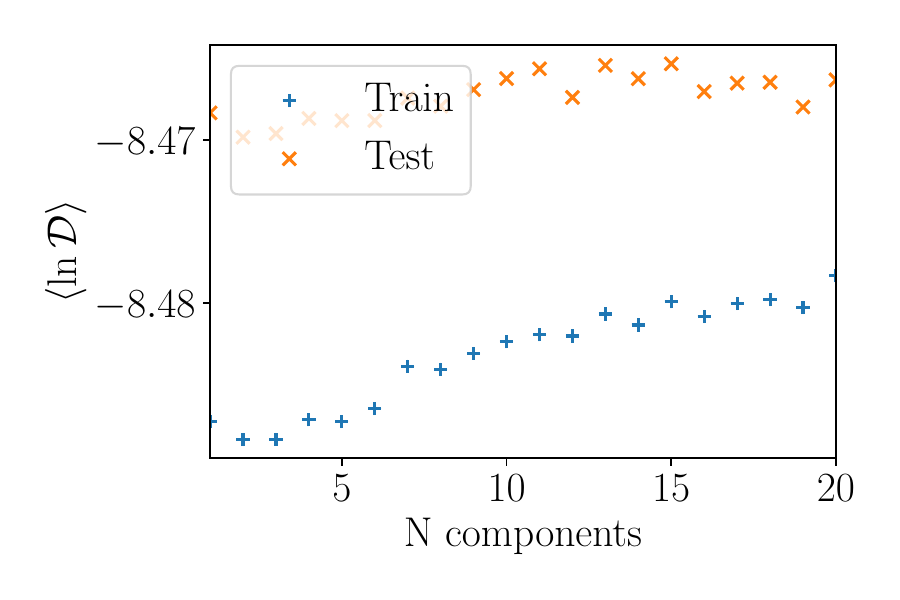}
    \caption{
    \revision{
    The average log-likelihood over the test and training data of the trained Gaussian mixture model as a function of the number of Gaussian components.
    We note that the performance on the test data set flattens out after $\sim 10$ components, while the performance on the training data continues to improve.
    }
    }
    \label{fig:components}
\end{figure}

\revision{
Our aim is to take the regularized samples $\{\theta'_{i}\}$ and estimate the density ${\cal D}$ using a Gaussian mixture model.
Training is performed by maximizing the mean natural log-density of the test samples as implemented in \textsc{scikit-learn}~\citep{scikit-learn}.
Adding more components will improve the quality of the fit.
However, by using too many components we risk over-fitting statistical fluctuations in the training set.
To avoid this, we split the samples into a training (80\%) and a test (20\%) set.
The fluctuations in the test set should be independent of those in the training set and therefore we will choose the number of components when the quality of the fit in the test set stops improving when adding more components.

In Figure~\ref{fig:components}, we show the average log-density over the training and test sample sets for the trained Gaussian mixture models with varying numbers of components.
The offset between the two sets of points is simply due to random fluctuations in the train/test split.
Other realizations can lead to a smaller mean log-density for the test data set.
We note that the performance on the test data set flattens out after $\sim 10$ components, while the performance on the training data continues to improve.
We, therefore, use $10$ components in the remainder of this work to avoid overfitting.
In the subsequent sections, we use a 10-component Gaussian mixture model density estimate ${\cal D}$ trained using all of the found injections $\{\theta'_{i}\}$.
}

\subsection{Evaluation}

After training the Gaussian mixture model, we can trivially generate new samples from the target distribution by drawing samples $\theta'$ from the Gaussian mixture model and applying the inverse of Eq.~\ref{eq:regularisation}.
Alternatively, since these are density estimates, we can also directly evaluate the estimated density
\begin{equation}
    \mathcal{D}(\theta') = \frac{p_{\text{det}}(\theta) p(\theta | \Lambda_0)}{ \mathcal{J}(\theta)}.
\end{equation}
In practice, we want to evaluate the selection function
\begin{equation}
    p_{\text{det}}(\theta) = \frac{\mathcal{J}(\theta) \mathcal{D}(\theta')}{p(\theta | \Lambda_{0})}.
\label{eq:pdet_density}
\end{equation}
Here $\mathcal{J}$ is the Jacobian from Equation~\ref{eq:jacobian} and $p(\theta | \Lambda_{0})$ is the original distribution of injections.
We use Eq.~\ref{eq:pdet_density} as an alternate means of computing Eq.~\ref{eq:Pdet} with an equivalent Monte Carlo integral over samples from the population distribution
\begin{equation}
    P_{\text{det}}(\Lambda) = \left< p_{\text{det}}(\theta_i) \right>_{\theta_i \sim p(\theta | \Lambda)}.
    \label{eq:Pdet_density}
\end{equation}
\revision{With the empirical mapping described in Section~\ref{sec:preprocessing} we have
\begin{equation}
    p_{\text{det}}(\theta) = \frac{\mathcal{D}(\theta')}{\mathcal{N}(\theta';\mu=0,\sigma=1)}
    \frac{\tilde{p}(\theta)}{p(\theta | \Lambda_{0})}.
\label{eq:pdet_density_sensible}
\end{equation}
This can be very efficiently evaluated as required.
}

We note that this method requires an efficient method of generating samples from the population distribution.
This can be trivially performed using inverse-transform sampling if the population model has an analytically invertible cumulative distribution function, or is a sum of such distributions.
For other population models \revision{(for example}, those using the low-mass smoothing introduced in~\cite{Talbot2018a} or the redshift distribution introduced in~\cite{Fishbach2018} and used in many population analyses, e.g.,~\citep{O2Pop, O3aPop}\revision{)} another method is required\revision{.
The simplest alternative approach is numerically estimating the inverse cumulative distribution function; this can be implemented for all models at an increased, and perhaps prohibitive, computational cost.
In~\cite{Wong2020a}, the authors train a deep flow-based generative network capable of very efficiently generating samples from the mass model in~\cite{Talbot2018a}.
Additionally, one could relegate the generation of samples from the population model to an offline pre-processing step by training a deep neural network to estimate $\Pdet$ directly.
}

In practice, we find that this method requires far fewer samples in the Monte Carlo integral than when resampling the found injections, $5000-10000$ samples from the population model versus $\sim 80000$ found injections with the same number of effective samples for each method.
The number of effective samples is defined slightly differently for the two Monte Carlo methods considered here.
For Equation~\ref{eq:Pdet_density} we adopt the usual definition~\citep{Elvira2018}
\begin{equation}
    N_{\text{eff}} = \frac{\left(\sum_{i=1}^{N} \pdet(\theta_i)\right)^2}{\sum_{i=1}^{N} \pdet(\theta_i)^2}.
\label{eq:neff}
\end{equation}
However, for Equation~\ref{eq:Pdetmc} a correction is required to account for the initial injections with $\pdet=0$~\citep{Farr2019}.
Following~\cite{Farr2019}, for both of these methods we only allow for samples with $N_{\text{eff}} > 4 N_{\text{events}}$ and marginalize over the statistical uncertainty in $\Pdet$ in the likelihood.

\section{\label{sec:results}Results}

\revision{
To demonstrate the efficacy of our new method we consider the accuracy and precision of three different methods to compute the population selection function
We evaluate $\Pdet$ three times for each this set of samples.
\begin{enumerate}
    \item Using Equation~\ref{eq:Pdetmc} with the original found injection set.
    \item Using Equation~\ref{eq:Pdetmc} with samples generated using our Gaussian mixture model.
    \item Using Equation~\ref{eq:Pdet_density} with \revision{10000} samples from the population model.
\end{enumerate}

We compare our estimators using 5000 samples drawn from the prior distribution for our population parameters specified in Table~\ref{tab:prior}.
These are the parameters describing the models presented in Section~\ref{sec:models}.
We note that the prior is specified on the mean and variance of the Beta distribution
\begin{align}
    \mu_{\chi} &= \frac{\alpha_{\chi}}{\alpha_{\chi} + \beta_{\chi}} \\
    \sigma^{2}_{\chi} &= \frac{\alpha_{\chi} \beta_{\chi}}{\left(\alpha_{\chi} + \beta_{\chi}\right)^2\left(\alpha_{\chi} + \beta_{\chi} + 1\right)} \\
    \alpha_{\chi} &= \frac{\mu^2_{\chi} (1 - \mu_{\chi}) - \mu_{\chi} \sigma^2_{\chi}}{\sigma^2_{\chi}} \\
    \beta_{\chi} &= \frac{\mu_{\chi} (1 - \mu_{\chi})^2 - (1 - \mu_{\chi}) \sigma_{\chi}^2}{\sigma^2_{\chi}}
\end{align}

We highlight one specific limiting case that will most clearly demonstrate the differences between the methods.
We note that $\alpha_{\chi} \rightarrow 0$ indicates a spin distribution that peaks very sharply at zero black hole spin.
If $\alpha_{\chi} < 1$, the distribution is singular at $a=0$; due to Monte Carlo convergence issues, these singular configurations have not been used in many previous analyses (e.g.,~\cite{O2Pop, O3aPop, O3bPop}.)
}

\begin{table}[]
    \centering
    \begin{tabular}{c|c}
        Parameter & Distribution \\
        $\alpha$ & $\mathcal{U}(-2, 8)$ \\
        $\beta$ & $\mathcal{U}(-2, 8)$ \\
        $m_{\min}$ & $\mathcal{U}(2, 10)$ \\
        $m_{\max}$ & $\mathcal{U}(50, 100)$ \\
        $\lambda_{m}$ & $\mathcal{U}(0, 1)$ \\
        $\mu_{m}$ & $\mathcal{U}(20, 50)$ \\
        $\sigma_{m}$ & $\mathcal{U}(0, 10)$ \\
        $\mu_{\chi}$ & $\mathcal{U}(0, 1)$ \\
        $\sigma_{\chi}^2$ & $\mathcal{U}(0, 0.25)$ \\
        $\xi$ & $\mathcal{U}(0, 1)$ \\
        $\sigma_{s}$ & $\mathcal{U}(0, 4)$ \\
    \end{tabular}
    \caption{Prior distribution for population parameters used in the analysis presented here.
    $\mathcal{U}(a, b)$ indicates a uniform distribution in $[a, b]$.
    We note that the Beta distribution is parameterized in terms of the mean ($\mu_{\chi}$) and variance ($\sigma^{2}_{\chi}$).
    Additional cuts are imposed such that $\alpha_{\chi},\beta_{\chi}>0$.
    }
    \label{tab:prior}
\end{table}

\subsection{\label{sec:validation}Computing $\Pdet$}

\begin{figure}
    \centering
    \includegraphics[width=\linewidth]{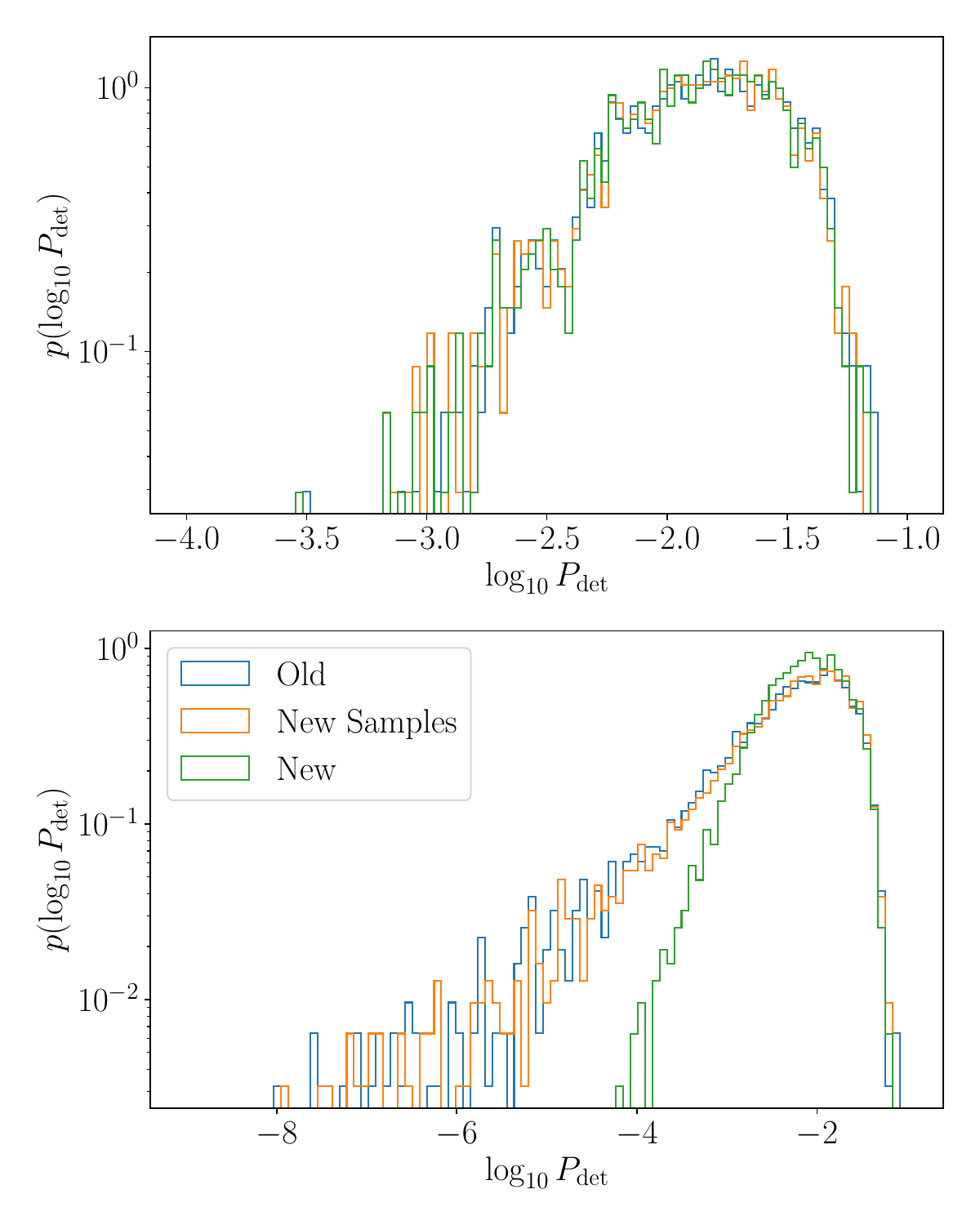}
    \caption{
    \revision{
    The distribution of the logarithm of the population-averaged sensitivity $\Pdet$ over the prior used for our training data (see Table~\ref{tab:prior}).
    In the top panel, we show samples for non-singular spin magnitude distributions ($\alpha_{\chi},\beta_{\chi}>1$).
    In the bottom panel, we use samples for singular spin distributions.
    In blue and orange, we calculate $\Pdet$ using Equation~\ref{eq:Pdetmc} as in previous analyses using the recovered simulated injections from the LIGO/Virgo collaboration~\cite{} (blue) and samples drawn from our Gaussian mixture model fit to $\pdet$ (orange).
    In green, we calculate $\Pdet$ using Equation~\ref{eq:Pdet_density} using 10000 samples from the population model.
    We note that the methods agree well for the non-singular distributions, however, the old method breaks down for singular spin distributions leading to the difference in calculated $P_{\rm det}.$
    }
    }
    \label{fig:Pdet}
\end{figure}

\begin{figure}
    \centering
    \includegraphics[width=\linewidth]{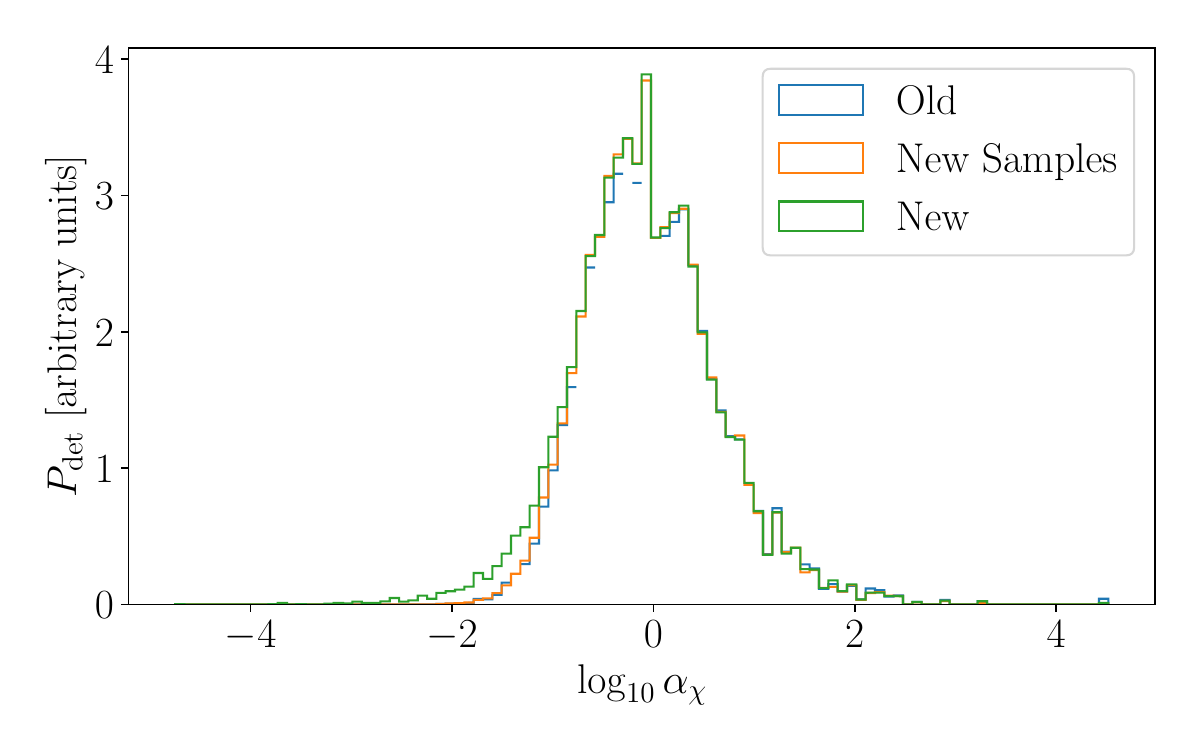}
    \caption{
    \revision{
    The population averaged sensitivity marginalized over all population parameters except the logarithm of the Beta distribution $\alpha$ parameter for spin magnitudes.
    In blue and orange, we calculate $\Pdet$ using Equation~\ref{eq:Pdetmc} as in previous analyses using the recovered simulated injections from the LIGO/Virgo collaboration~\cite{} (blue) and samples drawn from our Gaussian mixture model fit to $\pdet$ (orange).
    In green, we calculate $\Pdet$ using Equation~\ref{eq:Pdet_density} using 10000 samples from the population model.
    The computed values of $\Pdet$ agree for non-singular spin distributions $\log_{10}\alpha_{\chi} > 0$ but the sample reweighting method underestimates the sensitivity for singular spin configurations.
    }
    }
    \label{fig:Pdet_beta_sigma}
\end{figure}

While the mean log-density is a suitable metric for training our estimators, we can perform a stronger test by considering how well the population-averaged $\Pdet(\Lambda)$ compares across a range of $\Lambda$ values using our Gaussian mixture model.

\revision{In Figure~\ref{fig:Pdet}, we show the distribution $\Pdet$ for these three methods (in order blue, orange, green).
In the top and bottom panels, we show samples for non-singular ($\alpha_{\chi},\beta_{\chi}>1$) and singular spin magnitude distributions respectively.
We note that all the methods agree well for the non-singular distributions.
For singular spin distributions, both methods that resample a fiducial set of samples (blue and orange) methods differ significantly from our new method.
These distributions are very sharply peaked and hence are the natural failure mode of importance sampling.
The fact that the blue and orange results agree closely demonstrates that this effect is due to the difference in the construction of the Monte Carlo integral and not due to inaccuracy in the Gaussian mixture model estimate.

In Figure~\ref{fig:Pdet_beta_sigma}, we show the population-averaged sensitivity marginalized overall population parameters except for the logarithm of the Beta distribution $\alpha$ parameter.
The color scheme is the same as Figure~\ref{fig:Pdet}.
The fiducial sample reweighting methods underestimate the sensitivity to singular spin distributions compared to methods that directly evaluate $\pdet$.
We find that all the methods agree well for the parameters describing the black hole mass distribution.
}

\revision{
\subsection{\label{sec:mc-convergence}Monte Carlo Convergence}

As described in~\cite{Farr2019}, in order to have a reliable estimate of $\Pdet$, we need a sufficient number of effective samples in our Monte Carlo integral.
When reweighting a fixed set of recovered injections, this amounts to certain parts of the parameter space being inaccessible to our analyses.
By contrast, we can draw arbitrary numbers of samples from our density estimates of $\pdet$ in order to achieve sufficient convergence.

To demonstrate this, we compute the number of effective samples for each of the population samples used in the rest of this section using the Eqs~\ref{eq:Pdet} and~\ref{eq:Pdet_density}.
For the former, we use the $\approx 80000$ found injections during O3~\citep{O3bSensitivity}.
Throughout we use $N_{\rm events} = 69$ (BBH events confidently identified in O3.)
For the latter, we draw $10^{4}$ samples from the population model.
In Figure~\ref{fig:n_eff} we show the number of effective samples for the old fiducial sample reweighting method (blue with real injections and orange with samples drawn from our density estimate) and our new method (green) for two population parameters.
In the top and bottom panels, we show the parameters that demonstrate the most obvious trend for the old and new methods respectively ($\alpha_{\chi}$ and $\lambda_{m}$ the fraction of primary masses in the Gaussian component).
Once again, we note that the blue and orange and results agree closely, indicating the robustness of the Gaussian mixture model fit.

The spread in $N_{\rm eff}$ for the old method is over four orders of magnitude, while for the new method it is only two orders of magnitude.
This means that if using a fixed number of samples at each iteration, that fixed number can be smaller for the new method leading to increased computational efficiency.
As expected, we see that the efficiency of the fiducial sample reweighting drops significantly for $\alpha_{\chi} < 1$.
In contrast, our new method only weakly depends on the parameters of the spin magnitude distribution.
We find that when reweighting the found injections we reject $\approx 10\%$ of the non-singular samples and $\approx 50\%$ of the singular samples compared to $<1\%$ of samples using our new method.
The new method has more effective samples (smaller uncertainty in $>80\%$ of the space.)

\begin{figure}
    \centering
    \includegraphics[width=\linewidth]{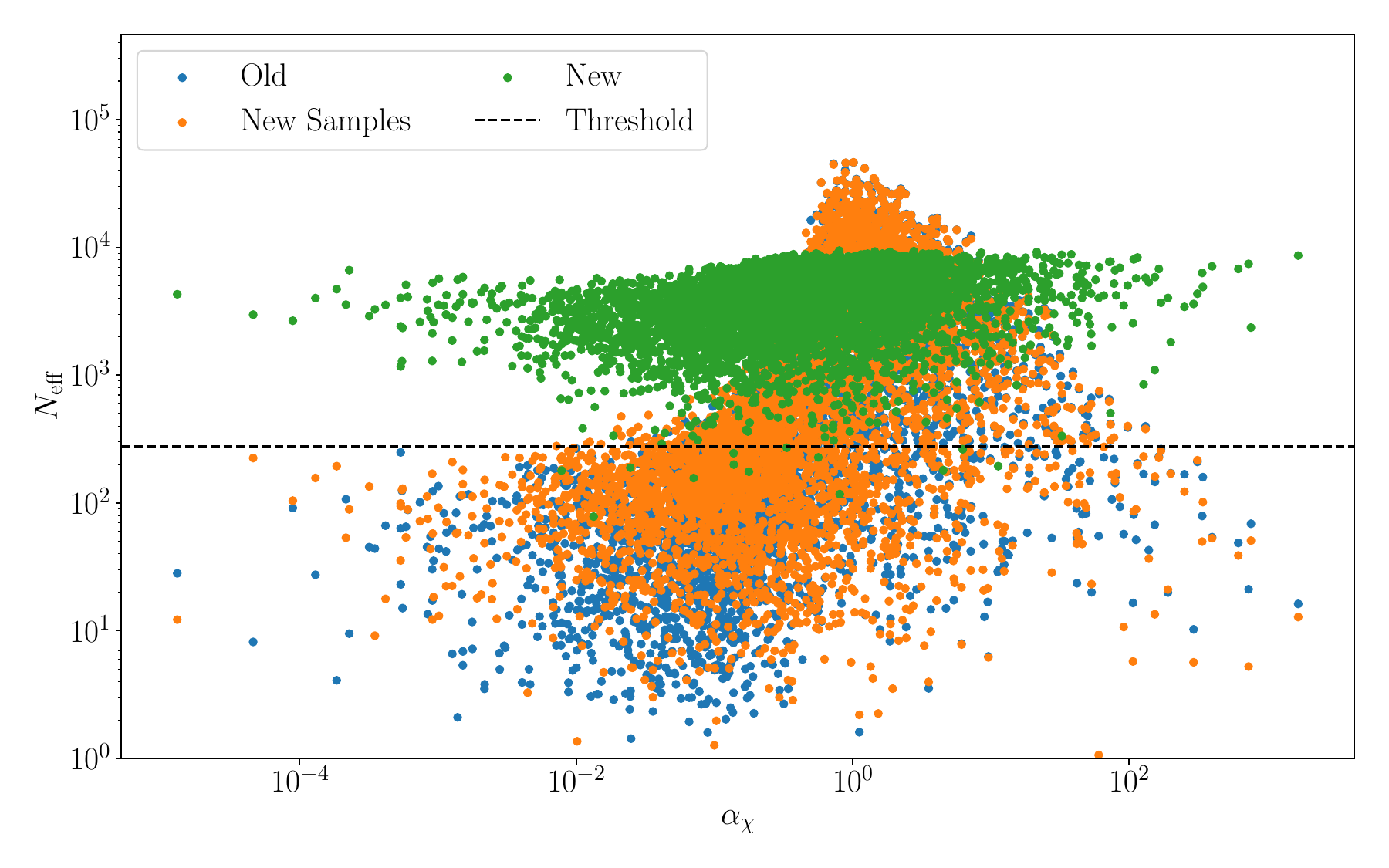}
    \includegraphics[width=\linewidth]{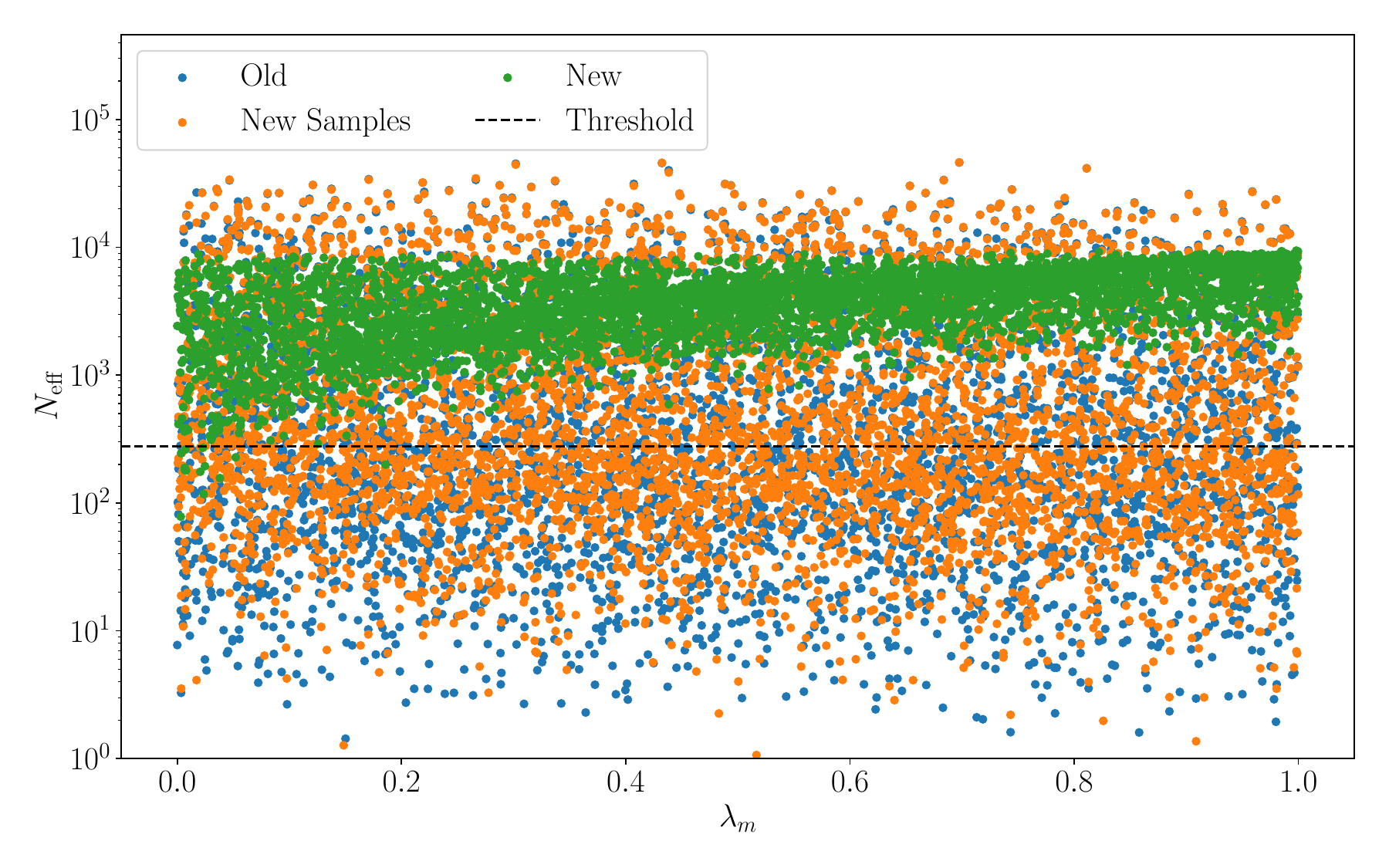}
    \caption{
    \revision{
        Number of effective samples in the calculation of $\Pdet$ as a function of two population parameters describing the distribution of merging black holes.
        (Top) One of the parameters describing the black hole spin distribution $\alpha_{\chi}$.
        (Bottom) The fraction of primary black holes whose mass falls in the Gaussian component of a two-component power-law and Gaussian mixture model.
        In blue, we show $N_{\rm eff}$ calculated by reweighting injections directly.
        In orange, we draw a new fixed set of samples using our Gaussian mixture model density estimate and reweight them.
        In green, we draw samples from the population model and directly evaluate $\pdet(\theta)$ using our Gaussian Mixture Model density estimate.
        The dashed black line shows the threshold for sufficient convergence of the Monte Carlo integral.
        We note that the old method is systematically biased away from small $\alpha_\chi$ (distributions that assign small spins to most black holes) and is unable to probe the region with $\alpha_{\chi} \lesssim 0.5$.
        In total, $\approx 40\%$ of the prior volume is inaccessible with the old method, compared to $< 1\%$ with our new method.
    }
    }
    \label{fig:n_eff}
\end{figure}

}

\subsection{\label{sec:hyperpe}Population inference}

\begin{figure}
    \centering
    \includegraphics[width=\linewidth]{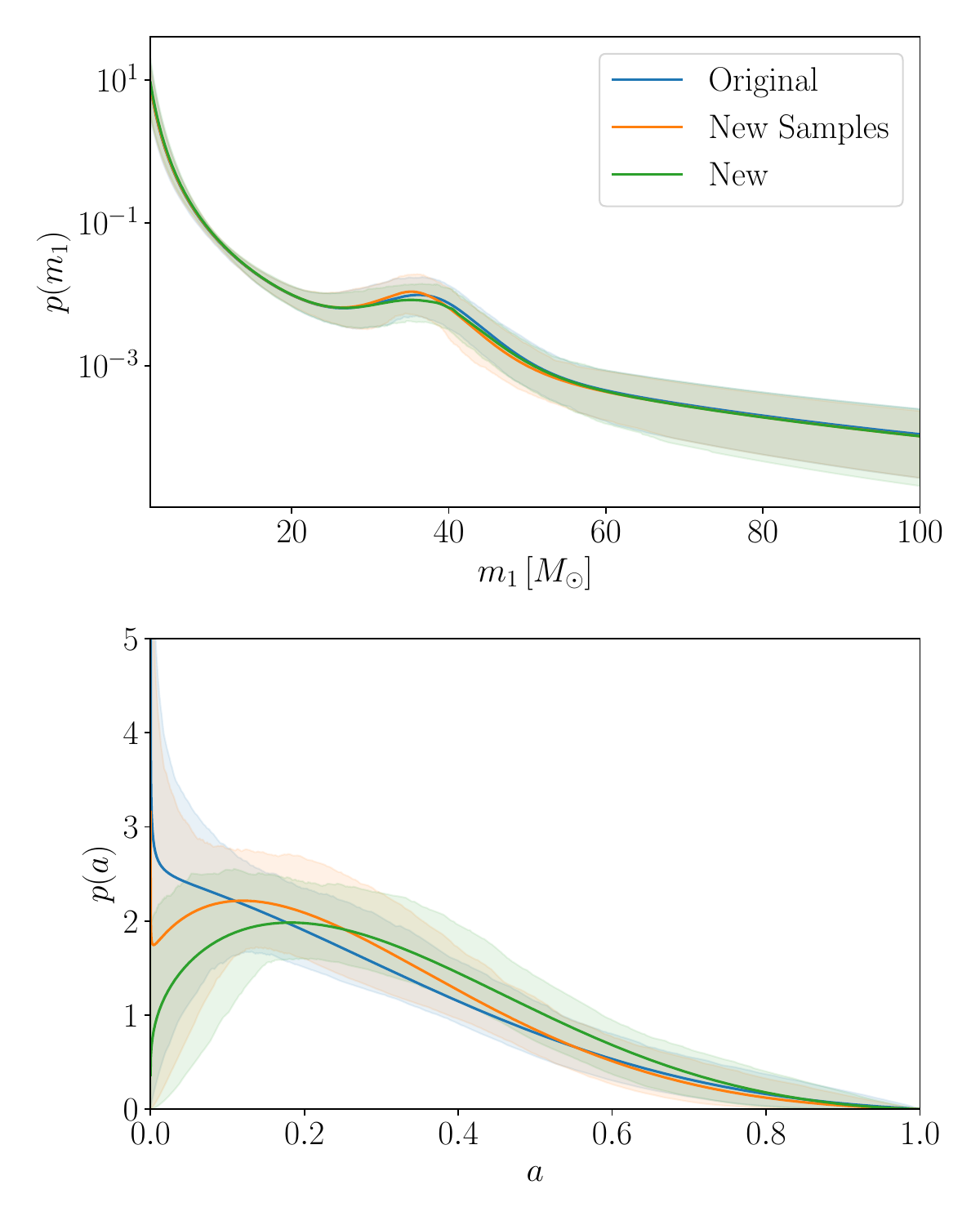}
    \caption{
    The inferred distribution of primary mass (top) and cosine spin tilt (bottom) when analyzing the binary black hole mergers identified in GWTC-3 using three different methods for estimating the \revision{population-averaged sensitivity}, $\Pdet$:
    (blue) using the original found injections, (orange) new samples generated using a Gaussian mixture model fit to the found injections, (green) evaluating $\pdet$ using samples from the population models.
    The results in blue and green do not include the dependence of $\Pdet$ on the spin distribution.
    We note that this leads to a slight change in the inferred distribution of spin orientations, when including spin dependence in the selection function there is slightly more preference for isotropic spins, although this is well below statistical uncertainties.
    The solid curves show the posterior predictive distribution and the shaded regions show the symmetric $90\%$ credible region.
    }
    \label{fig:hyperpe}
\end{figure}

We now consider the impact of our different $\Pdet$ evaluation methods on population inference.
\revision{We analyze the 69 binary black hole mergers with false alarm rate $<1~{\text{yr}^{-1}}$ identified in GWTC-3~\cite{GWTC3}.}
We perform population inference three times, once with each of our $\Pdet$ estimation methods using the \textsc{Bilby-MCMC} sampler~\citep{ashton2021}.

In Figure~\ref{fig:hyperpe}, we show the inferred astrophysical population of binary black holes when using our two methods to evaluate the selection function $P_{\text{det}}$.
In blue, we evaluate Equation~\ref{eq:Pdetmc} using the original simulated injections.
In orange, we evaluate Equation~\ref{eq:Pdetmc} using samples from our Gaussian mixture model.
In green, we evaluate Equation~\ref{eq:Pdet_density} using our Gaussian mixture model.
\revision{
The solid curves show the mean inferred distribution, while the shaded regions indicate the 90\% credible intervals.
We find no significant difference between the inferred population using these methods although we note a slight shift away from singular spin configurations when using our new method.
This is exactly the region in which the old method underestimates the sensitivity, leading to an overestimated likelihood.
}

\section{\label{sec:discussion}Discussion}

Malmquist biases are ubiquitous in astronomical surveys, and methods of understanding and mitigating these biases are vital to performing astrophysical inference.
Calculating and mitigating this bias is typically done by performing large simulations where synthetic signals are injected into the data and counting the number of recovered signals.
While this method gives an optimal estimate of the performance of searches for signals in real data, they can be difficult to work with and extend to generic population models.
\revision{In this work, we use these recovered signals to train a density estimate that can be reused to more efficiently compute the sensitivity to arbitrary populations.}

To \revision{improve the accuracy of} this density estimate we introduce a pre-processing step that improves the convergence.
Using this density estimate, we tested three methods to compute the population-level selection function.
We found that our new density estimation method matches the previous injection resampling method for population models \revision{where the Monte Carlo integrals are well converged.
We further demonstrated that our method is able to probe sharply peaked black hole spin distributions far more precisely than the existing method.}
This method can be more computationally expensive, especially for complex population models; however deep learning surrogate models present a solution to this problem~\cite{Wong2020a}.

Using our method, it is trivial to compute the fraction of sources which are observed $\Pdet$ by marginalizing over parameters other than those parameterized in the population model, e.g., evaluate $\Pdet$ using the parameters which most directly affect the sensitivity (chirp mass, mass ratio, effective aligned/precessing spin) and model the population in terms of parameters with the most intuitive physical meaning (component masses, spin magnitudes, and orientations).
We leave a detailed analysis of the best combination of parameters to use for the density estimation to future work.
Our results are consistent with the results presented in~\cite{O3bPop}; however, the uncertainty on the measured selection function is less in $>80\%$ of the space when using our new method.
As the catalog of observed compact binary coalescences grows it will be vital to understand the systematic error in our estimation of the selection function.

Machine learning methods for density estimation are rapidly gaining popularity in the gravitational-wave data analysis community, e.g.,~\cite{Powell2019, Gabbard2019, Green2020a, Green2021, Wong2021a, Wong2020a, Wong2020b, Cuoco2020}.
Most of these methods require the use of complex neural network-based density estimators which require tuning many more free parameters and thus extremely large training data sets.
The pre-processing method introduced here removes sharp spectral features, e.g., at prior boundaries, and thus enables high-precision estimation of the target distribution using Gaussian mixture models, rather than having to employ deep learning density estimators.
Combining this pre-processing in other density estimation problems may have a similarly simplifying effect.

One limitation of the current method is that the Gaussian mixture model employed in this work provides only a best-fit model and does not provide an indication of uncertainty in the fit over the parameter space.
We leave the exploration of density estimation techniques that model this uncertainty, e.g, Bayesian Gaussian mixture models to a future study.

\acknowledgments{
We thank Maya Fishbach for producing mock injections used in an early version of this work.
We thank Sylvia Biscoveanu, Tom Dent, Reed Essick, Jacob Golomb, Cody Messick, Richard O'Shaughnessy, Alan Weinstein, Daniel Wysocki, and Salvatore Vitale for useful comments and discussions.
This work is supported through the Australian Research Council (ARC) Centre of Excellence CE170100004 and ARC Future Fellowship FT150100281.
This is document LIGO-P2000505.
This research has made use of data, software, and/or web tools obtained from the Gravitational Wave Open Science Center (https://www.gw-openscience.org), a service of LIGO Laboratory, the LIGO Scientific Collaboration and the Virgo Collaboration.
Computing was performed computing clusters at the California Institute of Technology (LIGO Laboratory) supported by National Science Foundation Grants PHY-0757058 and PHY-0823459 and Swinburne University of Technology (OzSTAR).
This work used publicly available samples from~\cite{O2SampleRelease, O3aSampleRelease, O3aPopulationRelease, O3bSamples, O3bSensitivity}.
A {\textsc jupyter} notebook to fully reproduce the results presented here along with a number of additional diagnostic figures can be found on \href{https://github.com/ColmTalbot/gmm_sensitivity_estimation/blob/main/gmm_sensitivity_estimation.ipynb}{Github}.
This work made use of Google Colaboratory.
Software used in this work includes: {\sc numpy}~\citep{numpy}, {\sc scipy}~\citep{scipy}, {\sc scikit-learn}~\citep{scikit-learn}, {\sc matplotlib}~\citep{matplotlib}, {\sc pandas}~\citep{pandas}, {\sc cupy}~\citep{cupy}, {\sc Bilby}~\citep{bilby, ashton2021}, {\sc GWPopulation}~\citep{Talbot2019}
}

\bibliography{refs}

\end{document}